\documentclass[aps,preprint,showpacs]{revtex4}

\usepackage{amsmath}
\usepackage{graphicx}

\begin{document}

\title{Dynamics of broadband dispersive Alfv\'en waves}

\author{M.\ Marklund and P.K. Shukla}

\affiliation{Department of Physics, Ume{\aa} University, SE--901 87 Ume{\aa},
  Sweden}
  
\date{\today}
  
\begin{abstract}
  The properties of amplitude modulated broadband Alfv\'en waves is 
  investigated. In particular, the dynamics of circularly polarized dispersive Alfv\'en 
  waves, governed by a derivative nonlinear Schr\"odinger equation, is analyzed
  using the Wigner formalism. The modulational instability of random phase dispersive pump Alfv\'en waves is investigated, and it is shown that the spectral broadening gives rise to a new mode structure.  
\end{abstract}

\maketitle

\section{Introduction}

About thirty years ago, Rogister \cite{Rogister} introduced an elegant perturbation technique of the Vlasov--Maxwell system  of equations in order to investigate amplitude modulation of magnetic field-aligned circularly polarized dispersive Alfv\'en waves in an electron--ion plasma. He derived a derivative nonlinear Schr\"odinger (DNLS) equation, which is exactly integrable by the inverse scattering transform method \cite{Kaup-Newell}. The DNLS equation for nonlinear dispersive Alfv\'en waves have also been derived by means of the two-fluid -- Maxwell system of equations \cite{Mjolhus,Lashkin,Mio-etal,Spangler-Sheerin,Sakai-Sonnerup,Ruderman}, and the modulational instability of a constant a amplitude Alfv\'en pump wave has been investigated . Also reported is the algebraic envelope Alfv\'en soliton \cite{Mjolhus,Hada-etal}.

The DNLS equation has been extended to include the effects of collisions and wave-particle interactions \cite{Medvedev-Diamond,Medvedev-etal}. The resulting modified DNLS equation has been used to investigate damping of the envelope Alfv\'enic soliton. 

In this paper, we report on the nonlinear stability of broadband dispersive Alfv\'en waves in plasmas. We introduce a Wigner--Moyal formalism \cite{Wigner,Moyal} for the modified DNLS equation and derive a kinetic equation for Alfv\'enic quasiparticles. While the DNLS equation is appropriate for studies of modulational instability of a coherent Alfv\'en pump wave, the kinetic equation can be adopted to investigate the effects of partial coherence and random phases associated with broadband dispersive Alfv\'en wave packets.

\section{Governing equations}

Let us consider the nonlinear propagation of circularly polarized dispersive Alfv\'en waves along the external magnetic field $B_0\hat{\bf z}$, where $B_0$ is the magnetic field strength and $\hat{\bf z}$ is the unit vector along the $z$-axis. The dynamics of modulated Alfv\'en wave envelopes is governed by the modified DNLS equation \cite{Medvedev-Diamond,Medvedev-etal} 
\begin{equation}\label{eq:dnlse}
i\partial_tu + \partial_z^2u + i s\partial_z(|u|^2u) = p[u,u^*],
\end{equation}
where the subscripts $t$ and $z$ denotes the corresponding derivatives. 
Equation (\ref{eq:dnlse}) governs the amplitude modulation of circularly polarized Alfv\'en
waves.
Here $u=(B_x + i B_y)/(2|1-\beta|^{1/2}B_0)$,  $\beta=V_s^2/V_A^2$, $s=\text{sign}(1-\beta)$, time and space coordinates are normalized by the ion gyroperiod $\omega_{ci}^{-1}$ and the ion skin depth $V_A/\omega_{ci} =c/\omega_{pi}$. For our purposes, we  have in the collision-dominated case
\begin{equation}\label{eq:collision}
p=iD\partial_z^2u,
\end{equation}
where $D=(1/2)[(\eta_1/\rho_0) +c^2\eta_2/4\pi]\omega_{ci}/V_A^2$,
$\eta_1$ is the coefficient of the ion viscosity and $\eta_2$ is the
resistivity. On the other hand, in a collisionless case, we have
\begin{equation}\label{eq:collisionless}
p = i \frac{C}{4\pi} \partial_z\left(u H(z,t)\right)= 
i \frac{C}{4\pi} \partial_z\left(u \,\mathrm{P}\!\int_{-\infty}^{+\infty}dz^\prime \frac{|u(z^\prime,t)|^2}{z^\prime-z}\right),
\end{equation}
where $\mathrm{P}$ denotes the principal value. The coefficient $C$
depends on the velocity distribution of the particle species.
For $V_A \gg V_s$, we have
\begin{equation}
C=\sqrt{\frac{m_e}{2\pi m_i}}\,\left(\frac{V_s}{V_A}\right)\exp\left(-\frac{V_A^2}{2V_{Te}^2} 
\right) \ll 1,
\end{equation}
where $V_{Te}$ is the electron thermal speed.

With (\ref{eq:collision}) and (\ref{eq:collisionless}), Eq.\ (\ref{eq:dnlse}) is a 
special case of the equation
\begin{equation}\label{eq:gen-dnlse}
  i\partial_tu + g\,\partial_z^2u + F(|u|^2; t, z)u + i\partial_z[G(|u|^2; t, z)u] = 0 ,
\end{equation}
where $g$ is a complex valued constant and $F$ and $G$ 
are complex valued functions. This equation can be found in a wide
variety of applications, such as Bose-Einstein condensation and nonlinear
optics, apart from in plasma physics. Next we proceed to analyze Eq.\ 
(\ref{eq:gen-dnlse}) for broad band waves. 

In order to study the stability of broad band waves governed by 
Eq.\ (\ref{eq:gen-dnlse}), it is convenient to 
introduce the Wigner function 
\begin{equation}\label{eq:wignerfunc}
  \rho(t,z,k) = \frac{1}{2\pi} \int_{-\infty}^{\infty}\,d\zeta\,e^{ik\zeta}\langle 
    u^*(t, z + \zeta/2)u(t, z - \zeta/2) 
  \rangle
\end{equation}
for the Alfv\'en waves. Here the angular brackets represents the ensemble 
average. The Wigner function is given by the 
Fourier transform of the two-point correlation function, and 
as a generalized distribution function for quasi-particles it
can be used to describe a broad band spectrum of the field represented 
by $u$, e.g.\ Alfv\'en waves.
We note that from (\ref{eq:wignerfunc}) the equality
\begin{equation}
  I(t,z) \equiv \langle |u(t,z)|^2 \rangle = \int_{-\infty}^{\infty}\,dk\,\rho(t,z,k) 
\end{equation}
holds. 
Applying the time derivative to the Wigner function (\ref{eq:wignerfunc})
and using (\ref{eq:gen-dnlse}), one obtains the kinetic equation
\begin{eqnarray}
  &&
  \partial_t\rho + 2\,\mathrm{Re}(g)k\partial_z\rho 
  + \mathrm{Im}(g)\left(\tfrac{1}{2}\partial_z^2 - 2k^2\right)\rho
  + 2\,\mathrm{Im}\left[ F\exp\left( \tfrac{i}{2}\stackrel{\leftarrow}{\partial_z}
    \stackrel{\rightarrow}{\partial_k}\right) \right]\rho
\nonumber \\ &&\qquad
  + \partial_z\left\{ \mathrm{Re}\left[ G\exp\left( 
    \tfrac{i}{2}\stackrel{\leftarrow}{\partial_z}\stackrel{\rightarrow}{\partial_k}\right) 
  \right]\rho\right\}
  - 2k\,\mathrm{Im}\left[ G\exp\left( 
    \tfrac{i}{2}\stackrel{\leftarrow}{\partial_z}\stackrel{\rightarrow}{\partial_k}\right)
  \right]\rho
= 0 .
\label{eq:gen-kinetic}
\end{eqnarray}
As an example of the versatility of Eq.\ (\ref{eq:gen-kinetic}), we look at
the case $g = 1$, $F = I$, and $G = 0$. Then Eq.\ (\ref{eq:gen-dnlse}) reduces to the 
regular NLSE, and Eq.\ (\ref{eq:gen-kinetic}) becomes
\begin{equation}
  \partial_t\rho + 2k\partial_z\rho + 2I\sin\left( \tfrac{i}{2}\stackrel{\leftarrow}{\partial_z}
    \stackrel{\rightarrow}{\partial_k}\right)\rho = 0,
\end{equation}
an equation which to lowest order gives the Vlasov equation
\begin{equation}
  \partial_t\rho + 2k\partial_z\rho + (\partial_zI)(\partial_k\rho) = 0 .
\end{equation}

For Eq.\ (\ref{eq:dnlse}) we have $F = 0$, $g = 1 - iD$, and $G = sI - (C/4\pi)H$, 
and the kinetic equation (\ref{eq:gen-kinetic}) takes the form
\begin{eqnarray}
  \partial_t\rho + 2k\partial_z\rho - 2sk I \sin\left(  \tfrac{1}{2}\stackrel{\leftarrow}{\partial_z}
    \stackrel{\rightarrow}{\partial_k}\right)\rho
  + s\partial_z\left[ I \cos\left(
    \tfrac{1}{2}\stackrel{\leftarrow}{\partial_z}
    \stackrel{\rightarrow}{\partial_k}\right) \rho \right] 
    = \hat{L}\rho ,
\label{eq:alfven-kinetic}
\end{eqnarray}
where the operator expression on the right hand side is defined
according to 
\begin{subequations}
\begin{equation}
   \hat{L}\rho =  D\left(\tfrac{1}{2}\partial_z^2 - 2k^2\right)\rho 
\label{eq:collision-op}
\end{equation}
in the collisional case (\ref{eq:collision}), and 
\begin{eqnarray}
  &&
  \hat{L}\rho = 
      {-\frac{C}{2\pi}k\left[\mathrm{Im}(H)%
      \cos\left(  \tfrac{1}{2}\stackrel{\leftarrow}{\partial_z}
    \stackrel{\rightarrow}{\partial_k}\right)
      + \mathrm{Re}(H)%
      \sin\left(  \tfrac{1}{2}\stackrel{\leftarrow}{\partial_z}
    \stackrel{\rightarrow}{\partial_k}\right)\right]\rho}
\nonumber \\ &&\qquad 
    + \frac{C}{4\pi} \partial_z\left[ -\,\mathrm{Im}(H) \sin\left(
    \tfrac{1}{2}\stackrel{\leftarrow}{\partial_z}
    \stackrel{\rightarrow}{\partial_k}\right) \rho
      + \mathrm{Re}(H) \cos\left(
    \tfrac{1}{2}\stackrel{\leftarrow}{\partial_z}
    \stackrel{\rightarrow}{\partial_k}\right) \rho
    \right]
\label{eq:collisionless-op}
\end{eqnarray}
\end{subequations}
in the collisionless case (\ref{eq:collisionless}).

\section{Analysis of the coherent case}

We note that Eq.\ (\ref{eq:dnlse}) has solutions of the modified plane wave form 
$u(t,z) = \bar{u}(t)\exp[ik_0z - i\omega_0(t) t]$. In the collisional case 
(\ref{eq:collision}) the time-dependent frequency is given by  
\begin{subequations}
\begin{equation}
  \omega_0(t) = k_0^2 + \frac{su_0^2}{2Dk_0t}[1- \exp(-2Dk_0^2t)] , 
\end{equation}
while the amplitude is exponentially damped according to 
\begin{equation} \label{eq:collision-damped}
  \bar{u}(t) = u_0\exp(-Dk_0^2t) .
\end{equation}
\end{subequations}
Here $u_0$ is the constant amplitude of the solution, and we note that as $D \rightarrow 0$
we obtain the time-independent dispersion relation $\omega_0 = k_0^2 + sk_0u_0^2$. In 
the collisionless case (\ref{eq:collisionless}) however, the character of the function $p[u,u^*]$ 
together with the above plane wave ansatz makes the amplitude time-independent, and 
subsequently the frequency becomes 
\begin{equation}
  \omega_0 = k_0^2 + sk_0u_0^2 ,
\end{equation}
where $\bar{u} = u_0$ denotes the constant amplitude.
Thus, with the plane wave ansatz, the effects due to the thermal correction (\ref{eq:collisionless})
vanish.

Coherent modulational instabilities can be analyzed by letting 
$u = (\bar{u}(t) + u_1(t,z))\exp[ik_0z - i\omega_0(t)t] + \mathrm{c.c.}$, where 
$|u_1| \ll \bar{u}$, the time variation of $\bar{u}$ is assumed slow compared
to the time scale of the perturbation, and $\mathrm{c.c.}$ denotes the complex conjugate. Linearizing Eq.\ (\ref{eq:dnlse}) with 
respect to $u_1$, dividing $u_1$ into its real and imaginary part, and 
assuming the wavenumber and frequency $K$ and $\Omega$ of the perturbation
we find the dispersion relation
\begin{subequations}
\begin{equation}\label{eq:dispersion-collision}
  \Omega = -iDK^2 + 2k_0K + 2sK\bar{u}^2 \pm \left[ 
    s^2K^2\bar{u}^4 + K^4 + 2sk_0K^2\bar{u}^2  
    - 4Dk_0K(Dk_0K + iK^2 + isk_0\bar{u}^2)
  \right]^{1/2}
\end{equation}
in the collisional case (\ref{eq:collision}), and 
\begin{equation}\label{eq:dispersion-collisionless}
  \Omega = - \tfrac{1}{4}iCK\bar{u}^2 + 2k_0K + 2sK\bar{u}^2  \pm \left[ 
    s^2K^2\bar{u}^4 + K^4 + 2sk_0K^2\bar{u}^2
    - \tfrac{1}{2}CK^2\bar{u}^2\left( 
      \tfrac{1}{8}C\bar{u}^2 + ik_0 + is\bar{u}^2 
    \right)  
  \right]^{1/2}
\end{equation}
\label{eq:dispersion-mono}
\end{subequations}
in the collisionless case (\ref{eq:collisionless}). We note from 
(\ref{eq:dispersion-collision}) and (\ref{eq:dispersion-collisionless}) that for $s = 1$ 
there is no instability and the perturbation modes are always damped. 
However, for $s = -1$ this is not the case. 
Letting $\gamma = \mathrm{Im}(\Omega)$ where $\gamma$
is the growth rate, we find from (\ref{eq:dispersion-collision})
\begin{equation}\label{eq:growth-collision}
  \gamma = -DK^2 + \left( 
    2k_0K^2\bar{u}^2  - K^2\bar{u}^4 - K^4 
  \right)^{1/2} ,
\end{equation}
if we assume that $D \ll 1$ and linearize for $D$.
Similarly, in the collisionless case, the parameter $C \ll 1$, and we may expand
(\ref{eq:dispersion-collisionless}), keeping terms linear in $C$.
Then we find
\begin{equation}\label{eq:growth-collisionless}
  \gamma = - \tfrac{1}{4}CK\bar{u}^2 + \left(
     2k_0K^2\bar{u}^2 - K^2\bar{u}^4 - K^4    
  \right)^{1/2} .
\end{equation}
Thus, the expressions (\ref{eq:growth-collision}) and (\ref{eq:growth-collisionless})
show that for $s = -1 \Leftrightarrow V_s > V_A$ we have growing modes due to
the nonlinear evolution of the Alfv\'en waves. However, these modes 
suffer damping due to dissipative processes. 

Next we will compare the above results for the coherent modulational instability 
with the case of partial coherent Alfv\'en waves.

\section{Effects of partial coherence}
 
Returning to Eq.\ (\ref{eq:alfven-kinetic}), we first look for solutions 
of the form $\rho = \bar{\rho}(t,k)$. In the collisional case (\ref{eq:collision-op})
Eq.\ (\ref{eq:alfven-kinetic}) takes the form
$
  \partial_t\bar{\rho} + 2Dk^2\bar{\rho} = 0.
$ 
 This is integrated to yield 
\begin{equation}\label{eq:collision-distribution}
  \bar{\rho} = \rho_0(k)\exp(-2Dk^2t) .
\end{equation}  
Equation (\ref{eq:collision-distribution}) is in agreement 
with the monochromatic result (\ref{eq:collision-damped}) based on Eq.\ (\ref{eq:dnlse}),
if we let $\rho_0(k) = I_0\delta(k - k_0)$. In the collisionless case 
(\ref{eq:collisionless}), the ansatz for $\rho$ gives $H = 0$, where 
$\bar{I}(t) = \int dk\,\bar{\rho}(t,k)$. Thus, the 
expression (\ref{eq:collisionless-op}) gives
$
  \bar{\rho}  = \rho_0(k)
$ 
 through Eq.\ (\ref{eq:alfven-kinetic}), in agreement with the monochromatic result.

Next we look at perturbation around the background solutions. Thus, we let 
$\rho = \bar{\rho}(t,k) + 
\rho_1(k)\exp(iKz - i\Omega t) + \mathrm{c.c.}$, where $|\rho_1| \ll \rho_0$ and 
$|\partial_t\bar{\rho}| \ll \Omega\bar{\rho}$, and linearize
with respect to $\rho_1$. With $I = \bar{I}(t) + I_1\exp(iKz - i\Omega t) + \mathrm{c.c.}$ 
it follows that $H \approx \bar{H} + H_1 = H_1 = i\pi I_1\exp(iKz - i\Omega t) 
+ \mathrm{c.c}$. Thus, using
\begin{subequations}
\begin{equation}
  2\sin\left( \frac{i}{2}K\partial_k\right)\bar{\rho}(t,k) = 
    i\left[\bar{\rho}(t, k + K/2) - \bar{\rho}(t, k - K/2) \right]  
\end{equation}
and
\begin{equation}
  2\cos\left( \frac{i}{2}K\partial_k\right)\bar{\rho}(t,k) = 
  \bar{\rho}(t, k + K/2) + \bar{\rho}(t, k - K/2) ,
\end{equation}
\end{subequations}
we find that the nonlinear dispersion relation
\begin{equation}\label{eq:dispersion-general}
  1 = \frac{(s - ia)}{2K}\int\,dk\frac{(k - K/2)\bar{\rho}(t, k + K/2) - (k + K/2)\bar{\rho}(t, k - K/2)}%
    {k + [sK\bar{I}(t) - \Omega  - iL(k)]/2K}
\end{equation}
valid for partially coherent Alfv\'en waves. Here $L(k) = D(K^2/2 + 2k^2)$ and 
$a = 0$ in the collisional
case (\ref{eq:collision}), and $L(k) = 0$ and $a = C/4$ in the collisionless case 
(\ref{eq:collisionless}).
In the monochromatic case, we have $\rho_0(k) = I_0\delta(k - k_0)$.
The dispersion relation (\ref{eq:dispersion-general}) then becomes
\begin{eqnarray}
  &&\!\!\!\!\!\!
  \Omega =  - \tfrac{1}{2}i(L_+ + L_-) + 2 k_0K  + (2s - ia)K\bar{I} 
\\ &&\!\!\!\!\!\! 
    \pm \left\{ (s - ia)^2K^2\bar{I}^2 + K^4 + 2(s - ia)k_0K^2\bar{I} 
      - (L_+ - L_-) \left[\tfrac{1}{4} (L_+ - L_-) + iK^2 + i(s - ia)k_0\bar{I}\right]\right\}^{1/2} 
\nonumber 
\end{eqnarray}
where $L_{\pm} = L(k_0 \pm K/2)$. Thus, inserting the definitions of $L(k)$ and $a$ we
regain the dispersion relations (\ref{eq:dispersion-mono}) when neglecting the slow 
background time variations.

\begin{figure}
\includegraphics[width=1.05\textwidth]{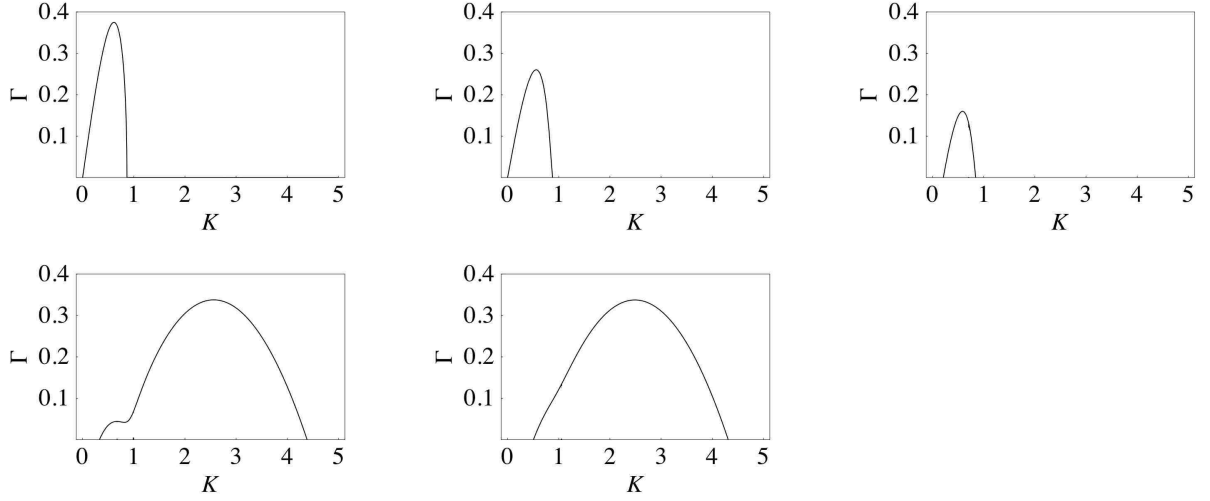}
\caption{The growth rate for different parameter values in the collisional case 
  (\ref{eq:collision}). The upper left panel has $s= -1$, while $D = \Delta = 0$; 
  the middle upper panel has $s= -1$ and $D = 0$, while $\Delta = 0.1$; 
  the right upper panel has $s = -1$ and $\Delta = 0$, while $D = 0.1$. 
  The left lower panel has $s = -1$ and $D = \Delta = 0.1$, while the right panel
  has $s = 1$ and $D = \Delta = 0.1$. Note that the last of these panels has no 
  coherent counterpart.}
\label{fig:1}
\end{figure}

However, if the Alfv\'en waves suffer random perturbations in e.g.\ the phase, the 
correlation between the waves may be nontrivial \cite{Klimontovich}. The partial coherence introduced by a random phase may be modeled by the Lorentzian distribution
\cite{Loudon}
\begin{equation}\label{eq:lorentz}
  \rho_0(k) = \frac{I_0}{\pi}\frac{\Delta}{(k - k_0)^2 + \Delta^2} ,
\end{equation}
where $\Delta$ denotes the width of the distribution around $k_0$. Due to the finite 
width of the Lorentz distribution the dispersion integral (\ref{eq:dispersion-general})
will have poles. In the collisional case (\ref{eq:collision}) we obtain a rather 
lengthy analytical expression, where a complicated interplay between the 
spectral broadening, represented by $\Delta$, and the collision parameter $D$
takes place. Thus, we do not explicitly state this particular result here, but instead give the numerical solution of (\ref{eq:dispersion-general}) in conjunction with the Lorentzian 
distribution (\ref{eq:lorentz}) in the collisional case (\ref{eq:collision}) 
(see Fig. \ref{fig:1}). However, in the collisionless case (\ref{eq:collisionless}), the result is somewhat more compact and takes the form 
\begin{eqnarray}
  &&\!\!\!\!\!\!\!\!\!
  1 = 2KI_0\left( s - \tfrac{1}{4}iC \right)\left\{\frac{ \bar{\Omega} + k_0K + iK\Delta}%
    {( \bar{\Omega} +  2iK\Delta )^2 - K^4} \right. 
\nonumber \\[2mm] &&\!\!\!\!\!\!\!\!\! \left.
  - iK{\Delta}\frac{{4 K^2\Delta^2 + (\bar{\Omega} - K^2 )^2/2 
    + (\bar{\Omega} + K^2)^2/2 + 2(\bar{\Omega} + 2k_0K)(\bar{\Omega} - sKI_0)}}{
{[4 K^2\Delta^2 + (\bar{\Omega} - K^2 )^2][4K^2\Delta^2 
+ (\bar{\Omega} + K^2 )^2]}} \right\} ,
\label{eq:gen-disp-collisionless}
\end{eqnarray}
where $\bar{\Omega} = \Omega - 2k_0K - sKI_0$.

\begin{figure}
\includegraphics[width=1.05\textwidth]{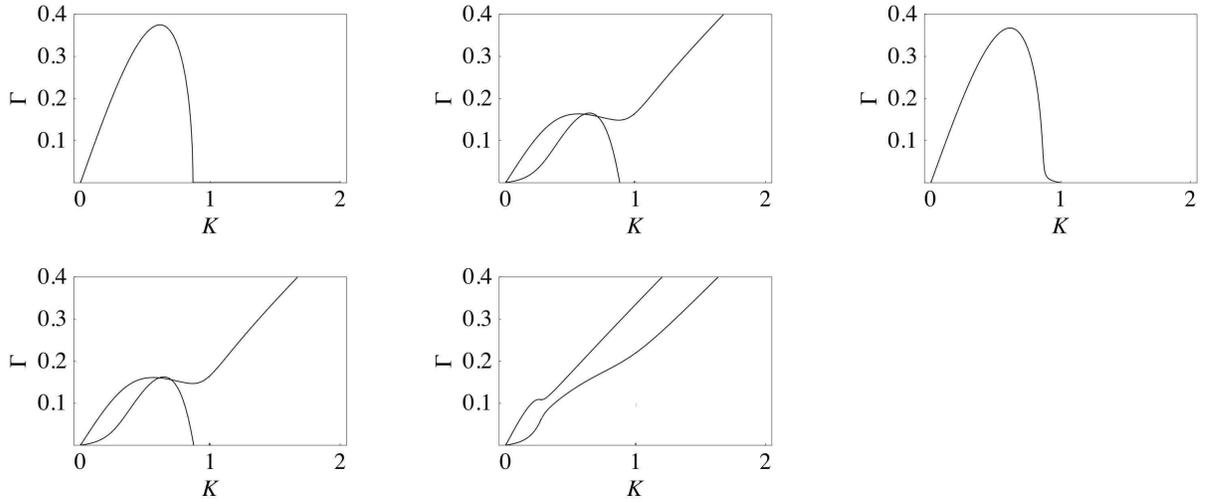}
\caption{The growth rate for different parameter values in the collisional case 
  (\ref{eq:collisionless}). The upper left panel has $s= -1$, while $C = \Delta = 0$; 
  the middle upper panel has $s= -1$ and $C = 0$, while $\Delta = 0.1$; 
  the right upper panel has $s = -1$ and $\Delta = 0$, while $C = 0.1$. 
  The left lower panel has $s = -1$ and $C = \Delta = 0.1$, while the right panel
  has $s = 1$ and $C = \Delta = 0.1$. Note that the last of these panels has no 
  coherent counterpart.}
\label{fig:2}
\end{figure}

Letting $\bar{\Omega} = \mathrm{Re}(\bar{\Omega}) + i\Gamma$, where $\mathrm{Re}$
denotes the real part, we can solve the nonlinear dispersion relations numerically for the growth rate $\Gamma$ as a function of the wavenumber $K$. The results are shown in Figs. \ref{fig:1} and \ref{fig:2}, where we have used $I_0 = 0.5$ and $k_0 = 1$. We see that the parameter region with $s = 1$ exhibits new properties as compared to the coherent case. A new type of instability occurs due to the spectral broadening of the dispersive Alfv\'en pump wave. 

\section{Summary}

We have investigated the nonlinear stability of broadband dispersive Alfv\'en waves in magnetoplasmas, using a Wigner--Moyal formalism. A new mode structure with new instabilities, due to the finite spectral width of the dispersive Alfv\'en pump wave, is found. The spectral broadening in conjunction with the kinetic modification of the DNLS equation can thus give rise to growing modes not present in the coherent case.

\newpage


\end{document}